\documentclass[3p,times]{elsarticle}

\usepackage{ecrc}
\usepackage{placeins} 

\volume{00}

\firstpage{1}

\journalname{Applied Soft Computing}

\runauth{}

\jid{procs}

\jnltitlelogo{-}

\CopyrightLine{2011}{Published by Elsevier Ltd.}

\usepackage{amssymb}

\usepackage[figuresright]{rotating}
\usepackage{listings}

\begin{document}

\begin{frontmatter}



\dochead{}


\title{Enhancing Phishing Detection in Financial Systems through NLP}


\author{Leminur Çelik}
\ead{celikl19@itu.edu.tr}
\author{Novruz Amirov}
\ead{amirov20@itu.edu.tr}
\author{Egemen Ali Caner}
\ead{caner19@itu.edu.tr}
\author{Emre Yurdakul}
\ead{yurdakule19@itu.edu.tr}
\author{Fahri Anıl Yerlikaya}
\ead{yerlikayaf@itu.edu.tr}
\author{Şerif Bahtiyar}
\ead{bahtiyars@itu.edu.tr}

\address{Department of Computer Engineering, Istanbul Technical University, Maslak, Istanbul 34469, Turkey}

\begin{abstract}

The threat of phishing attacks in financial systems is continuously growing. Therefore, protecting sensitive information from unauthorized access is paramount. This paper discusses the critical need for robust email phishing detection. Several existing methods, including blacklists and whitelists, play a crucial role in detecting phishing attempts. Nevertheless, these methods possess inherent limitations, emphasizing the need for the development of a more advanced solution. Our proposed solution presents a pioneering Natural Language Processing (NLP) approach for phishing email detection. Leveraging semantic similarity and TF-IDF (Term Frequency-Inverse Document Frequency) analysis, our solution identifies keywords in phishing emails, subsequently evaluating the semantic similarities with a dedicated phishing dataset, ultimately contributing to the enhancement of cybersecurity and NLP domains through a robust solution for detecting phishing threats in financial systems. Experimental results show the accuracy of our phishing detection method can reach 79.8 percent according to TF-IDF analysis, while it can reach 67.2 percent according to semantic analysis.

\end{abstract}


\end{frontmatter}


\section{Introduction}

\begin{figure}[h!]
  \centering
  \includegraphics[width=7cm,height=5.3cm]{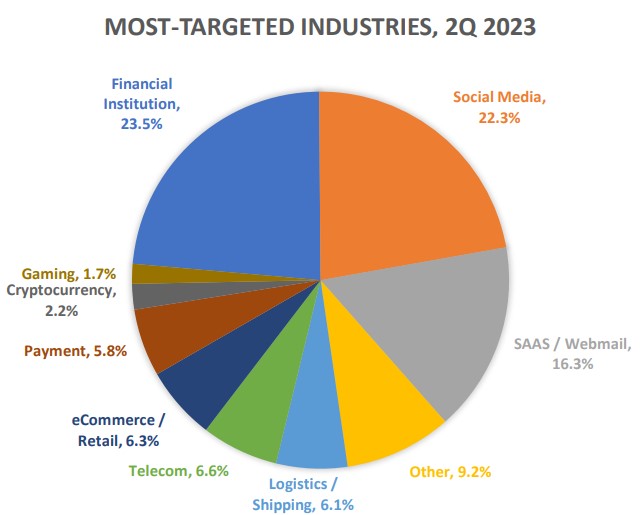}
      \caption{Most targeted industry sectors by phishing attacks \cite{apwg-report}.}
  \label{fig-most-targeted-industries}
\end{figure}

Phishing is a type of cyber-crime where a person pretending to be a member of a legitimate organization contacts a target via email, phone, or text message in an attempt to trick the target to reveal sensitive information, such as passwords, banking and credit card information, personally identifiable information, and etc. Many people are unable to recognize phishing emails and they share personal information with the attacker. Therefore, it becomes very significant to detect and prevent phishing emails to protect the society. On the other hand, phishing attacks grow continuously targeting specific sectors. In the second quarter of 2023, 1,286,208 phishing attacks were recorded by APWG between April and June 2023 \cite{apwg-report}. The APWG's third-highest quarterly total to date was recorded in the 2Q 2023. According to APWG, phishing attacks targeting the financial sector, which includes banks, remained the most prevalent kind of attacks in the second quarter of 2023, which accounts 23.5\% of all phishing attacks, as shown in Figure \ref{fig-most-targeted-industries}. Moreover, 6.3\% of all attacks targeted e-commerce and retail sector. In this research, we primary consider phishing attacks on financial systems.

\begin{figure}[h!]
  \centering
  \includegraphics[width=11cm,height=6cm]{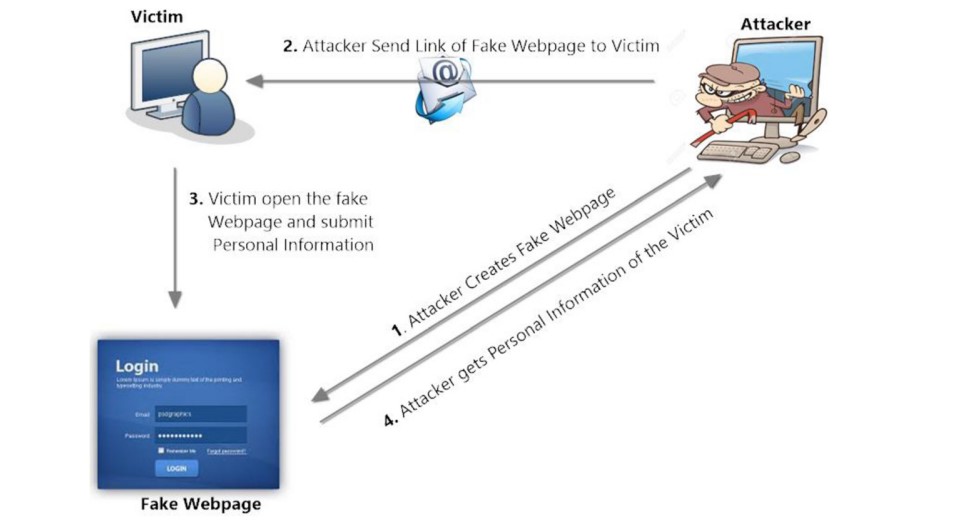}
      \caption{A phishing attacks life cycle \cite{phishing-life-cycle}.}
  \label{fig-phishing-life-cycle}
\end{figure}

A process explaining how phishing attacks occur is explained in Figure \ref{fig-phishing-life-cycle}. The initial stage in a phishing attacks is to create a visually convincing phishing website by copying content from a reputable business or bank's website. The attacker then creates and distributes emails with the link of the phishing website, either to a large audience or, in the case of spear phishing, to a specific target group. When a user opens the email and goes to the phishing website, they are asked to enter personal data. Exploiting the trust of individuals to reputable companies, especially when imitating the website of a bank, results in users inadvertently disclosing their credentials. After obtaining personal information, the attacker uses it for illegal or financial gain \cite{phishing-life-cycle}.

Since phishing attacks may have consequences, phishing detection is a significant research area. In general there are two approaches to detect phishing, namely list based and machine learning based. For instance, a blacklist contains malicious URLs, and whether the site to be entered is reliable or cannot be detected by comparing it with the URLs in the list. The whitelist includes safe sites, and if the site to be entered is not in this list, it is understood to be a phishing email. Machine learning models play an important role in detecting phishing emails. Phishing detection relies significantly on context-sensitive architecture of Natural Language Processing (NLP) that offers a remarkable ability to acquire, modify, and enhance their language production and comprehension abilities. Briefly, the two approaches are not mature enough to protect contemporary financial systems against phishing attacks. 

In this paper, we propose a novel approach for detecting phishing emails using NLP models. Our solution includes three phases. In the first phase, keywords of emails are identified by using semantic similarity and Term Frequency-Inverse Document Frequency (TF-IDF) analysis. In the second phase, following the acquisition of keywords, it is essential to assess the semantic similarities between keywords and a dataset comprising exclusively phishing emails. The final phase is the classification of emails as benign or phishing. The main contribution of our research is to propose a new approach that uses NLP with both semantic and non-semantic models.

The rest of the paper is organized as follows. Section \ref{sec:related-works} contains related works about phishing detection. Section \ref{sec:proposed-solution} presents our proposed solution. Section \ref{sec:analysis-solution} is about the analysis of our solution. We conclude the paper in Section \ref{sec:conclusion}.

\section{State of the Art about Phishing Detection}\label{sec:related-works}
The number of phishing attacks has increased dramatically due to high connectivity among people over the web. In this section, we analyze factors contributing to the increase and, more importantly, why the need for an effective defense mechanism against phishing attacks is urgent. 


The surge in phishing attacks can be attributed to the inherent vulnerabilities within email systems, which cybercriminals exploit using sophisticated tactics to impersonate trusted sources and compromise individuals' inboxes. Defense mechanisms must evolve to counter these threats and protect sensitive information. The State University of New York in Binghamton has highlighted various vulnerabilities in email systems over the years. Notable instances include session hijacking vulnerabilities in Gmail, Inbox Mail, and Hotmail in 2008, a cross-site scripting flaw in Gmail's iOS message attachment feature in 2013, and cross-site request forgery issues in Gmail's password reset system. Other vulnerabilities involve unauthorized access to contact lists in Gmail (2006), email and contact theft in Hotmail (2011), a successful breach of a customer's CloudFlare account using Google Apps (2012), and the compromise of over one billion Yahoo user accounts in 2013 \cite{security-analysis}.


In e-commerce, driven by widespread online shopping, has broadened the attack surface for cyber-criminals. The financial lure of phishing attacks on e-commerce platforms, fueled by increased online transactions, highlights the need for robust defenses. Evolving tactics of attackers in the fraud economy underscore the urgency for comprehensive security measures. The sector's reliance on user trust makes it susceptible to phishing, exploiting users' confidence through deceptive emails and websites \cite{ecommerce-trust-control}. Fraudulent transactions in e-commerce rise significantly within just two quarters, reaching almost two and a half times the total transactions, from 0.8\% to 2.1\%. Particularly, high-demand sectors like luxury goods face strategic targeting, emphasizing the need for online merchants to enhance security measures against evolving threats in the dynamic e-commerce landscape \cite{phishing-and-ecommerce}.



Skula and Kvet \cite{domain-blacklist} explained the blacklist creation method that describes how to find and classify phishing domains by creating a blacklist and graylist. Following a sequential check, the records were categorized as ambiguous (UNK), non-phishing (FP), or confirmed phishing (TP). A domain was classified as FP if it was on the graylist and as TP if it was on the blacklist. Only the records that updated the blacklist were kept after records not found in either list were updated according to the original source classification. A dataset of confirmed instances of phishing was produced, along with a Graylist of FP records and a Blacklist of TP records. This process makes sure that phishing domains were managed in a chronological order.


Sharifi and Siadati \cite{blacklist-generator} suggested a blacklist generator that takes a URL as input and determines if the website is legitimate or phishing. The method uses all pages with the same hostname to predict every website. Their algorithm makes use of phishing pages and datasets of authentic websites that are chosen through a nearly uniform web sampling technique.

Azeez et al. \cite{whitelist-approach} explained the whitelist that is established through a comparison between the real link and the visual representation, as well as by assessing similarities with known trusted websites. The proposed solution uses this information, which is obtained from user-supplied web addresses, to make final determinations.

According to the study of Cao, Han and Le \cite{automated-whitelist}, Automated Individual White-List (AIWL) tracks familiar Login User Interfaces (LUIs) on websites for every user using a whitelist. LUIs are legitimate login pages where users enter their username and password. An alert is raised, signaling a possible attack, if a user tries to enter sensitive data into a non-whitelisted, unknown LUI. A Naïve Bayesian classifier is used by AIWL to determine which login attempts are successful. A website becomes familiar to the user after several successful log-in attempts, at which point the user confirms the addition of the LUI information to the whitelist, thereby improving security.

\begin{figure*}[h!]
	\centering
	\includegraphics[width=\textwidth]{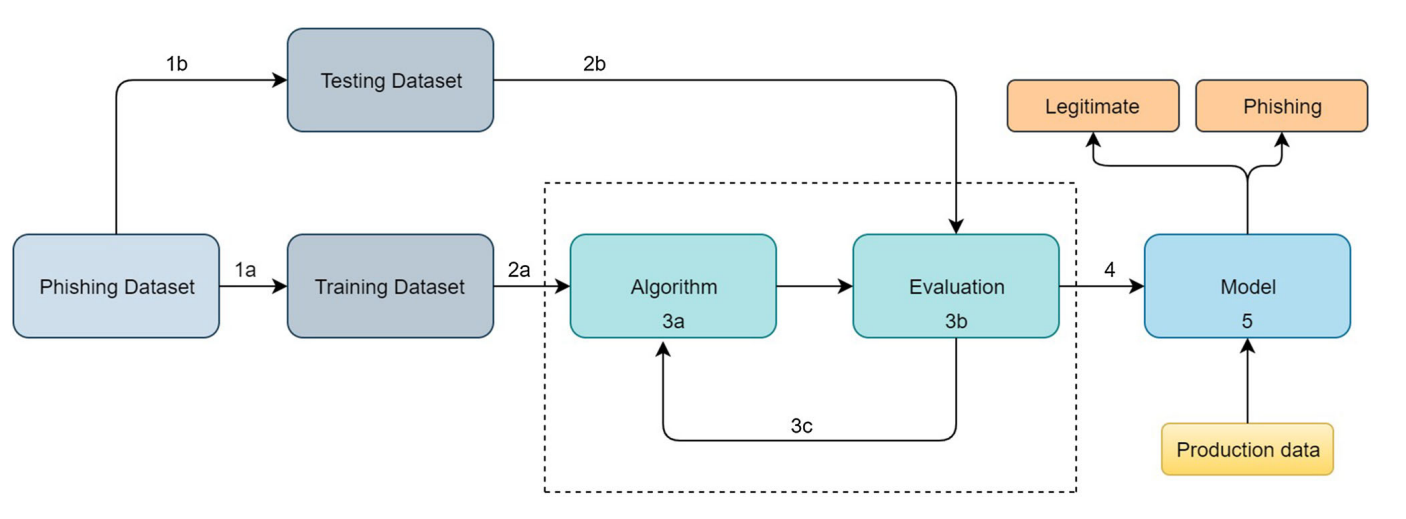}	
	\caption{Machine learning for phishing attack detection \cite{figure}.}
	\label{ml_phishing_detection}
\end{figure*}

Recently, machine learning (ML) has been used to detect phishing attacks. Figure \ref{ml_phishing_detection} illustrates the initial steps in employing machine learning for this purpose, emphasizing the importance of curating a phishing dataset, splitting it for adapting and testing, and optimizing the model's ability to discern phishing attempts. ML proves indispensable in countering the dynamic and sophisticated nature of phishing attacks \cite{ml_essential}.

Phishing attacks, growing in complexity, often evade conventional security measures. ML's adaptability and learning capabilities enable it to recognize new patterns and threats. ML models, continuously trained and analyzed, swiftly detect subtle variations and evolving tactics, serving as a defense against the ever-changing threat landscape \cite{ml_essential}. Notably, ML-based classifiers achieve accuracy rates exceeding 99\%, showcasing their efficacy in distinguishing between benign and malicious content \cite{ml_99}. This precision, a testament to ML's analytical prowess and adaptability, reinforces its pivotal role in cybersecurity defenses against phishing threats.

Ensemble ML techniques, exemplified by Random Forest (RF), emerge as powerful tools in phishing detection. Combining diverse ML models, these ensembles enhance prediction accuracy by reducing the risk of incorrect decisions. Their flexibility, adjusting weights during iterative procedures, renders them more reliable and stable, making them ideal for combating the dynamic nature of phishing attacks \cite{ml_ensemble}. Ensemble methods are ideal for countering the dynamic and ever-changing nature of phishing attacks because of their flexibility, which also makes them more stable and dependable than traditional machine learning algorithms.

Natural Language Processing (NLP) plays a pivotal role in various domains due to its evolving capabilities in understanding and processing human language. In the financial sector, NLP-based solutions surpass traditional static approaches for processing corporate financial reports. A study comparing syntactic features obtained through NLP with shallow readability formulas found that while syntactic features present challenges in computational complexity, NLP's nuanced understanding of readability differences across language varieties outperforms traditional methods \cite{nlp-shallow}. For phishing detection, NLP's context-sensitive structure is crucial. NLP's use of neural network-derived embedding allows it to excel in understanding word semantics within context. This sensitivity enables NLP to identify subtle linguistic cues and patterns in email content, making it a powerful tool for detecting phishing attempts that might elude conventional methods \cite{nlp_vector}. NLP's inherent adaptability is a significant advantage \cite{nlpfast}.

TextRank is a graph-based ranking algorithm for natural language processing tasks such as text summarizing. The algorithm works by representing sentences as nodes in a graph, with edges denoting relationships between sentences. By iteratively ranking and selecting sentences based on their importance in the graph, TextRank produces concise and informative document summaries \cite{nlp-rouge}. Thus, NLP's versatility and adaptability make it a powerful tool in a various areas of finance. It demonstrates its ability to automate tasks, enhance precision, and address challenges in dynamic and complex data environment, such as phishing detection.


To sum up, the distinction between machine learning and blacklist-based approaches in phishing detection is evident. Blacklist mechanisms rely on manual identification and reporting, demanding significant human resources and time. In contrast, machine learning methods automate identification using classification algorithms, excelling at feature extraction without labor-intensive manual engineering \cite{ml_blacklist_difference}. This highlights the efficiency of Machine Learning in addressing phishing threats compared to the potentially less effective blacklist mechanisms, especially in scenarios requiring timely detection \cite{blacklist-dl}.


Machine learning solutions outperform blacklisting due to implementation differences. While blacklisting struggles with evolving phishing attacks, machine learning dynamically learns from historical data, identifying patterns not explicitly listed in blacklists. List-based methods are fast but ineffective against zero-hour attacks. In contrast, machine learning detects on-the-fly and handles zero-hour attacks effectively \cite{zerohour-attack}.

Existing researches show that list based mechanisms are faster but they are unable to detect phishing attacks that are dynamic. On the other hand, machine learning mechanisms are slower but have potential to detect evolving phishing attacks. Furthermore, NLP based approaches are a good candidate to detect phishing attacks in financial systems.

\section{NLP based Phishing Detection Model for Financial Systems}\label{sec:proposed-solution}
We propose a new phishing detection model that uses a sequential approach leveraging NLP techniques. Initially, top keywords are extracted from a dataset of phishing emails. Subsequently, a semantic analysis is conducted to assess the similarity between these keywords and the content of the emails. To establish a discriminatory semantic threshold between phishing and legitimate emails, a separate dataset is utilized for testing purposes. Lastly, after obtaining the semantic threshold value and extracted words, the detector is ready to test email data.

\subsection{Phase 1: Extracting Words}
Keywords in phishing emails were determined with two different methods, namely semantic similarity method and TF-IDF method. The first method processes text data from a CSV file. It combines and pre-processes the text, extracting keywords through K-means clustering algorithm. The process involves tokenization, lemmatization, and removal of stopwords and punctuation. The output consists of top keywords representing the clustered word vectors. 

\begin{figure} [h!]
    \centering
    \includegraphics[width=1.0\textwidth]{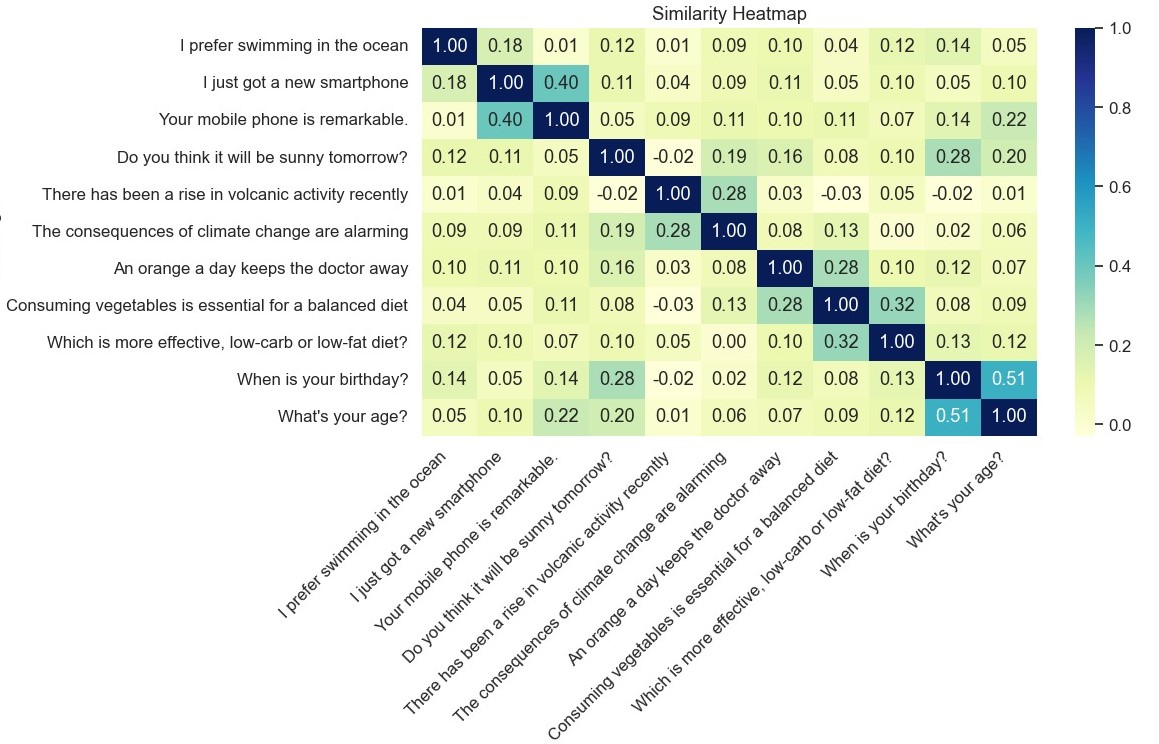}
    \caption{Semantic heat map.}
    \label{heat_map}
\end{figure}

The main logic of phase one involves reading a CSV file, combining all text from a specified column into a single document, and getting embedding. After obtaining vectors of each unique lemmatized word, K-means clustering is applied to the vectors. One variable controls the number of clusters. Next, by determining the indices and minimal distances between cluster centers and word vectors, the script determines which words from each cluster are the most representative. A list of the most used keywords that are obtained from the input text data is the final product. The code prints the most used keywords. As an example of semantic similarity, the heat map is shown in Figure \ref{heat_map}.

The second method uses TF-IDF representation to extract keywords from text data in a CSV file. The first step is to import libraries required for natural language processing, TF-IDF vectorization, and data manipulation. After that, tokenizing the input text, changing its case, eliminating stopwords, and keeping only alphanumeric words are the steps taken by the preprocessing.

\begin{figure}[h!]
  \centering
  \includegraphics[width=0.7\textwidth]{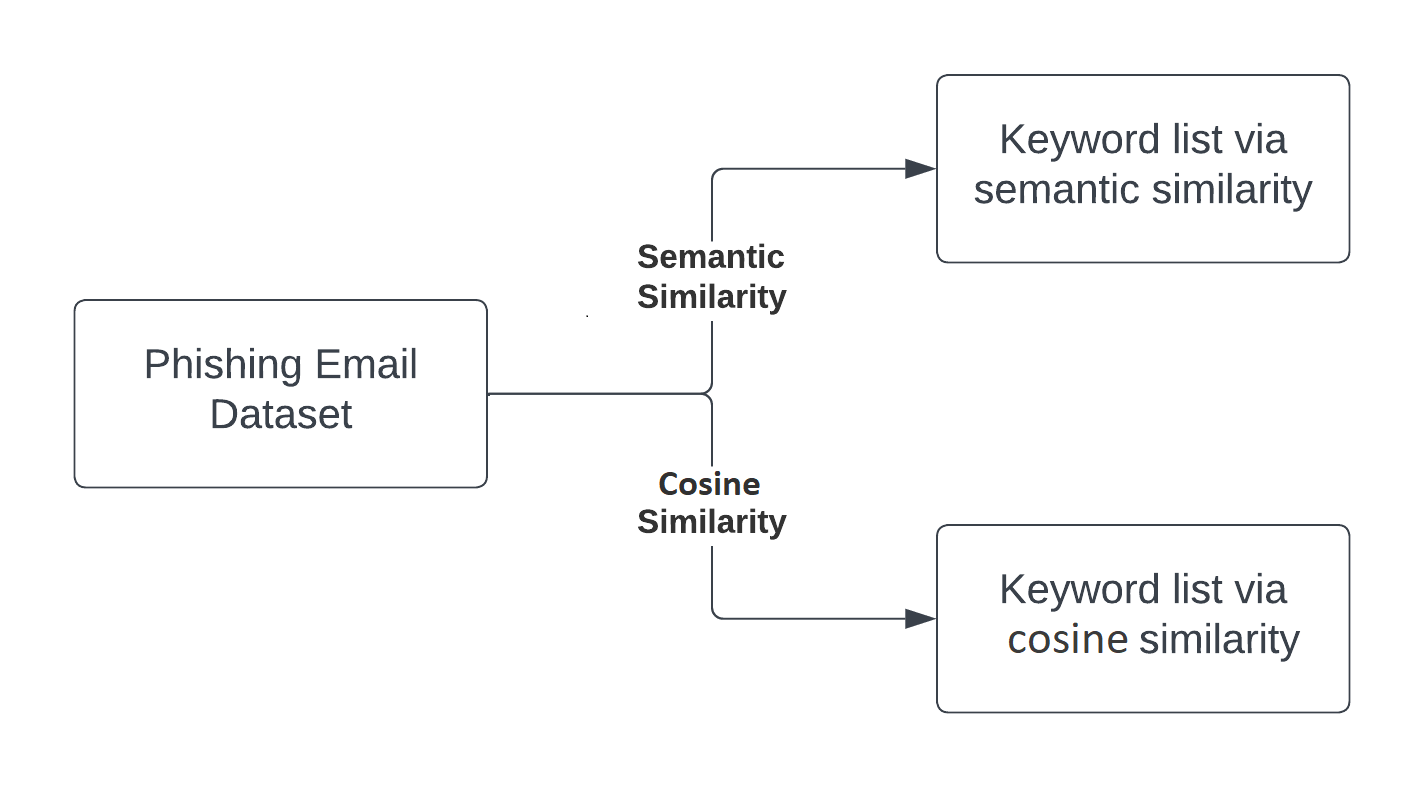}
  \caption{Define keyword list using similarity algorithms.}
  \label{fig-keywords}
\end{figure}

To assign importance scores to words, the main function reads CSV file, extracts text from a given column, combines and preprocesses the text, and then uses TF-IDF vectorization. It uses these TF-IDF scores to determine the most used N keywords and returns them. The most used keywords are printed out by applying this functionality to a particular CSV file and column. Once frequently occurring words in emails are identified, they are routed to Adapting phase as in Figure \ref{fig-keywords}.

\subsection{Phase 2: Adapting}
After obtaining the keywords, semantic similarities between the keywords and a dataset consisting of only phishing emails are examined. To find the semantic similarities, Universal Sentence Encoder (USE) \cite{USE}, a model designed to convert variable-length sentences into fixed-size embeddings, capturing semantic information effectively is used. The initial steps involve loading USE model and defining a function to embed input sentences. The cosine similarity between pairs of sentence embeddings is measured through dot product operations. The main logic then assesses semantic similarity by calculating the cosine similarity between each email text and keyword list. After analyzing the semantic similarities, a threshold value is defined  for the dataset and a keyword list by heuristically examining breakpoints about relation between threshold increase and success rate gain to determine whether the corresponding similarity refers to 'phishing email' or 'safe email'. Lastly, obtained threshold value and the pair of dataset and keyword list is redirected to 'testing phase'.

\subsection{Phase 3: Testing}
In the testing phase, the phishing detection model is tested by evaluating its performance on real-world email data. The goal is to effectively classify emails as either phishing or benign based on semantic similarities with the previously extracted keyword list. Steps are taking shaped from semantic analysis of data, adapting threshold selection, and model prediction.

The model utilizes a set semantic threshold to evaluate the similarity between the content of incoming emails and a predefined keyword list. By converting email content into fixed-size embeddings, the model captures underlying semantic information. The cosine similarity metric is then employed to measure the proximity of each email to the identified keywords.

\begin{figure}[h!]
  \centering
  \includegraphics[width=0.9\textwidth]{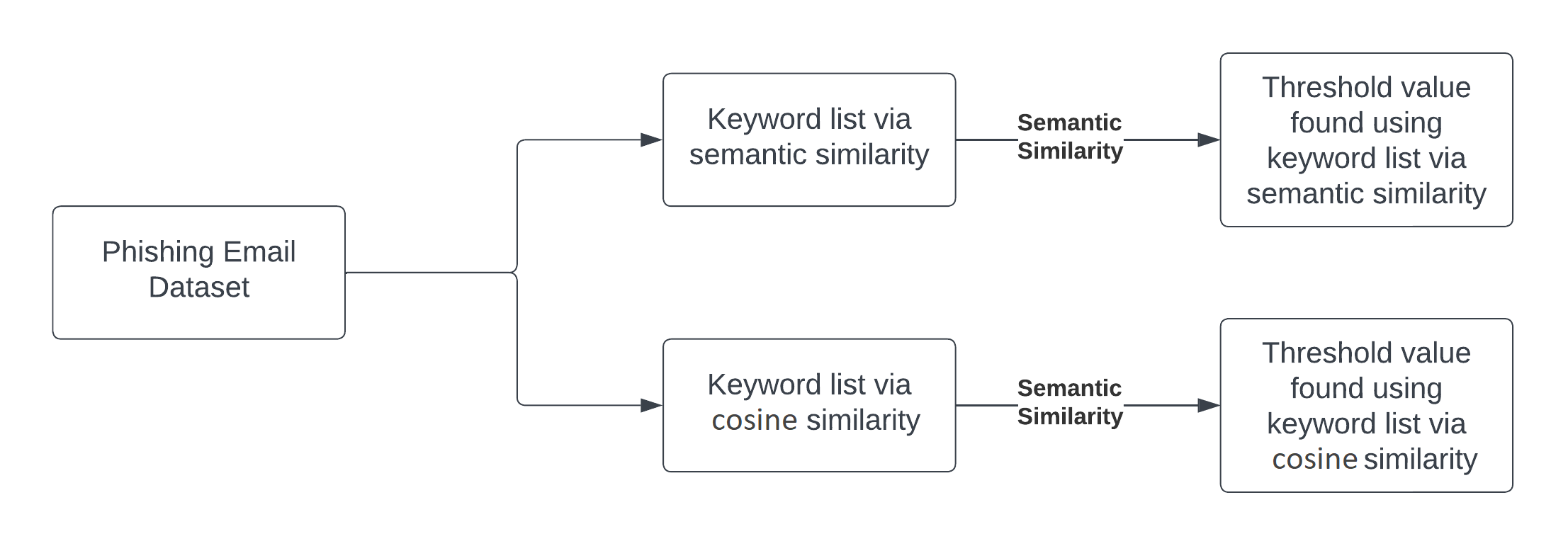}
  \caption{Find threshold values using keyword lists.}
  \label{fig-threshold}
\end{figure}

Recognizing the dynamic nature of phishing attacks, the system incorporates a mechanism for adaptive threshold adjustment that as in Figure \ref{fig-threshold}. Machine learning algorithms, such as reinforcement learning, is employed to continuously optimize the threshold based on the evolving characteristics of phishing emails, yet the threshold value is found heuristically by examining breakpoints.

\begin{figure}[h!]
  \centering
  \includegraphics[width=0.9\textwidth]{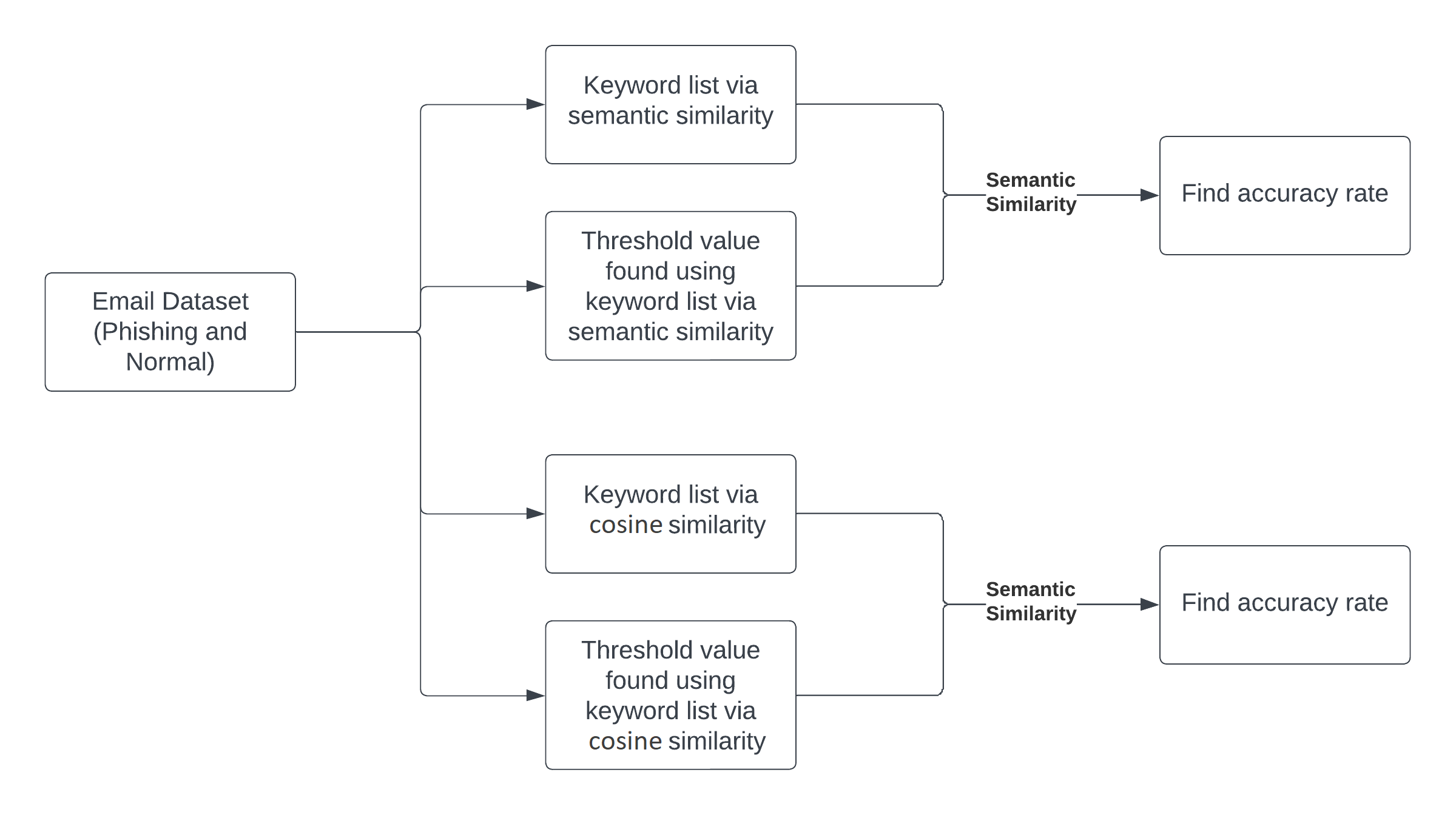}
  \caption{Find accuracy rates using keyword lists and threshold values.}
  \label{fig-accuracy}
\end{figure}

A decision threshold is used to convert confidence scores into actionable decisions. If the confidence score of an email exceeds the threshold, it is confidently classified as phishing email. Otherwise, the email is classified as safe, benign. The threshold acts as a decision point, enabling users to determine the classification outcome based on whether the score surpasses or falls below the determined threshold so that accuracy rates can be calculated as shown in Figure \ref{fig-accuracy}. This mechanism provides flexibility to fine-tuned balance between sensitivity and specificity in the email classification process.

\section{Analysis of Proposed Model}\label{sec:analysis-solution}

\subsection{Dataset Overview}
The details of datasets used for keyword extraction, threshold adapting, and comprehensive testing. A meticulous examination of these datasets lays the foundation for the subsequent in-depth analysis of both semantic and non-semantic approaches. The significance of dataset selection cannot be overstated, as it plays a pivotal role in shaping the models' adaptability and overall performance in practical scenarios.

The primary dataset, comprising 190 emails, serves as the bedrock for our keyword extraction process. Carefully curated from a publicly accessible source, this dataset has undergone rigorous selection to encompass a diverse and extensive array of phishing email scenarios. The deliberate selection ensures that both semantic and non-semantic models face a broad range of plausible threats \cite{extract-dataset}. Each email within this dataset contributes uniquely to the development and refinement of our models, enabling them to discern subtleties in phishing attempts with heightened accuracy.

Our second dataset, housing 250 phishing emails obtained from Kaggle, is a strategic choice aimed at running both the semantic and non-semantic approaches \cite{kaggle-dataset}. A renowned hub for machine learning datasets, this collection holds paramount importance in shaping the interpretative abilities of our model concerning various phishing patterns. The decision to incorporate a diverse and realistic component to the adapting process is intentional, leveraging Kaggle's reputation for providing high-quality datasets.

\begin{figure}[h!]
  \centering
  \includegraphics[width=0.7\textwidth]{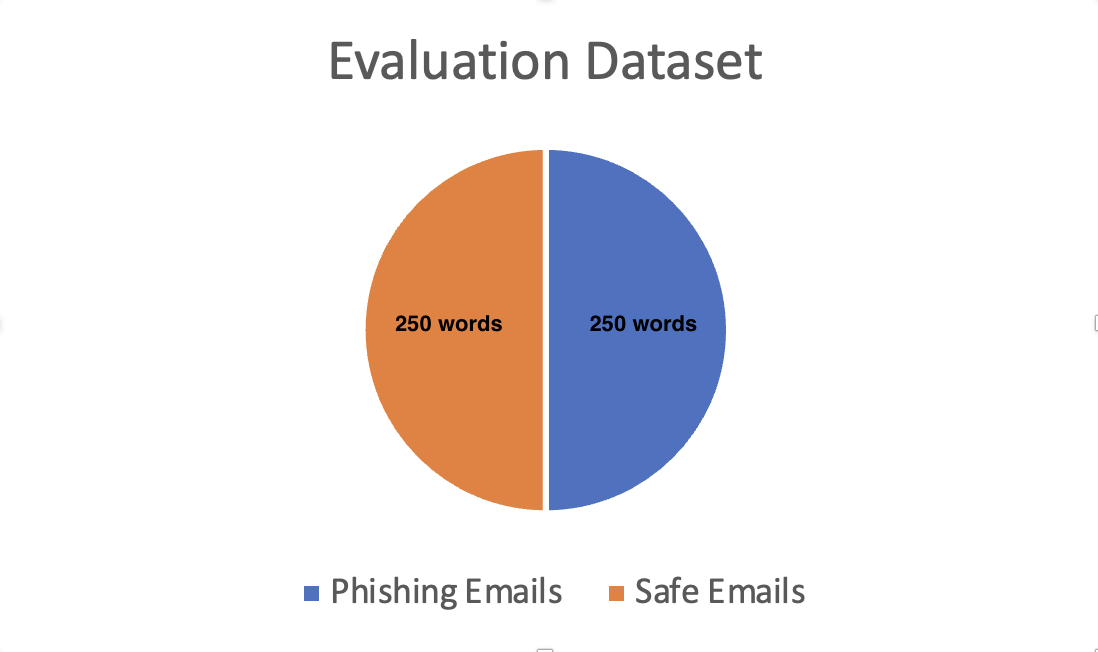}
  \caption{Dataset used for evaluation process of our model, 500 emails.}
  \label{fig-evaluation-dataset}
\end{figure}

A third dataset comprising 500 emails (250 phishing and 250 safe) for model testing is shown in Figure \ref{fig-evaluation-dataset}. The same dataset, which is the dataset for adapting the threshold, is used for testing. This dataset plays a crucial role in subjecting our model to a rigorous evaluation that mirrors the real-world challenge of differentiating between malicious and benign emails. The inclusion of equal proportions of phishing and safe emails in this dataset ensures a balanced assessment of our models' efficiency. For example, an email body with "Message sent trough eBay System Your registered name is included to show this message originated from eBay. Learn more. Account On-hold: Please confirm your eBay informations today" is phishing email. Whereas, email body with "Yes, very much so, Mi Vida. Thank you for thinking of me. Sent from my Wild iPhone" is safe email. Moreover, understanding the nuances embedded within each dataset is paramount for a holistic comprehension of our model's adaptability, and eventual performance in practical scenarios. 



\subsection{Semantic Model Evaluation}
\begin{figure}[h!]
  \centering
  \includegraphics[width=0.8\textwidth]{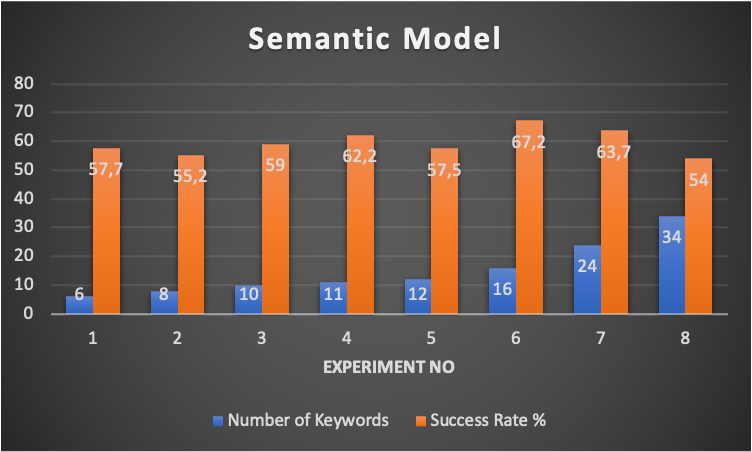}
  \caption{Output for semantic model.}
  \label{fig-semantic-model}
\end{figure}

The performance of the semantic model yields subtle insights from five tests. Results are shown in Figure \ref{fig-semantic-model}. Utilizing 6 keywords, experiment 1 establishes a foundational success 57.7\%, serving as a benchmark for subsequent assessments. Experiment 2, employing 8 keywords, demonstrates a slightly diminished success rate of 55.2\%. The introduction of 10 keywords in experiment 3 results in a notable increase in success rate to 59\%, signaling the presence of information saturation. The discriminatory capability of the model is compromised by the noise introduced through an excessive number of keywords.

Experiment 4, utilizing 11 keywords, showcases a moderate recovery in accuracy, achieving a success rate of 62.2\%. This emphasizes the delicate balancing act required when selecting keywords to ensure optimal performance. Experiment 5, featuring with 12 keywords, records a success rate of 57.5\%.

Expanding our exploration, experiment 6 introduces 16 keywords, yielding a success rate of 67.2\%. Experiment 7, with 24 keywords, attains a success rate of 63.7\%, while experiment 8, utilizing 34 keywords, achieves a success rate of 54\%. These experiments enrich our understanding of the semantic model's dynamics, offering insights into its adaptability and responsiveness to variations in the number of keywords.

The significance of optimizing the trade-off between comprehensiveness and relevance in keyword extraction is highlighted by the semantic model's sensitivity to the quantity of keywords. A greater number of keywords may be able to catch more subtle aspects, but doing so may increase noise, therefore the trade-off between sensitivity and specificity must be carefully considered. 

\subsection{TF-IDF Model Evaluation}
\begin{figure}[h!]
  \centering
  \includegraphics[width=0.8\textwidth]{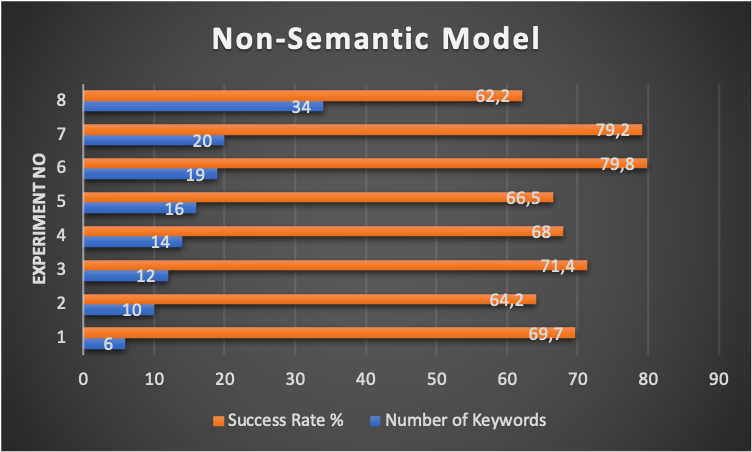}
  \caption{Output for non-semantic model.}
  \label{fig-non-semantic-model}
\end{figure}

The non-semantic model consistently exhibits robust performance across an extended set of eight experiments, as depicted in Figure \ref{fig-non-semantic-model}. The goal of this analysis is to uncover the enduring strength and adaptability of the non-semantic model for phishing email detection. Experiment 1, featuring 6 keywords, establishes an impressive success rate of 69.7\%, showcasing the model's proficiency. Experiment 2 with 10 keywords reporting a success rate of 64.2\%, underlining the model's resilience. Experiment 3 introduces a broader set of 12 keywords, achieving a commendable success rate of 71.4\%. This emphasizes the versatility of the model, suggesting its ability to adapt to a wider array of keywords from a diverse phishing dataset.

Experiment 4, utilizing 14 keywords, demonstrates the model's flexibility by maintaining a success rate of 68\%, even with variations in the quantity of keywords. Experiment 5, with 16 keywords, attains a success rate of 66.5\%, confirming the non-semantic model's consistent and reliable performance across varying keyword counts. Expanding our exploration, experiment 6 introduces 19 keywords, achieving a success rate of 79.8\%. Experiment 7, with 20 keywords, attains a success rate of 79.2\%, while experiment 8, featuring 34 keywords, reports a success rate of 62.2\%. These experiments further solidify the non-semantic model's position as a dependable and versatile option for phishing email detection.

The evaluations not only reinforce the non-semantic model's robustness but also shed light on adaptability to different keyword counts. The consistently high success rates across experiments underscore the reliability of the non-semantic approach, positioning it as a practical and resilient choice for detecting phishing threats.

The resilience of the non-semantic model is suggested by its ability to sustain high success rates across experiments, even with different keyword numbers, which is especially beneficial in real-world applications. The non-semantic method is positioned as a dependable and practical option for phishing email detection due to its ability to withstand variations in the quantity of keywords.

\subsection{Comparative Analysis}
The accuracy of the semantic model varies; it peaked with experiment 1 and then steadily decreased as the number of keywords increased. The non-semantic approach, on the other hand, continuously maintains high success rates, even when faced with a more extensive set of keywords. This disparity implies that, although the semantic model might perform better in specific scenarios, the non-semantic approach is a more reliable and flexible option for phishing email detection.

\section{Conclusion}\label{sec:conclusion}
Phishing attacks have significant consequences on financial systems. In this research, we evaluate two phishing detection by using NLP algorithms. We highlight the role of dataset selection that plays an important role in the performance of detection algorithms. Specifically, the semantic approach emphasizes the trade-off in keyword extraction even though it is sensitive to keyword quantity. On the other hand, the non-semantic model repeatedly demonstrates its robustness, making it a viable option for phishing email detection in a variety of datasets. 


As as future work, we decide to concentrate on optimizing the semantic model's keyword extraction procedure and investigating sophisticated methods in semantic analysis to augment discriminatory capability. 



\label{}





\bibliographystyle{elsarticle-num}







\end{document}